%% file: ndn-bootstrap.tex
\documentclass[sigconf, 10pt]{acmart}
\renewcommand\footnotetextcopyrightpermission[1]{} 
\usepackage{ifthen}
\usepackage{booktabs}
\usepackage{soul}
\usepackage{xspace}
\usepackage[normalem]{ulem}
\usepackage{graphicx}
\usepackage[inline]{enumitem}
\usepackage{etoolbox}
\usepackage{xstring}
\usepackage{tikz}
\usepackage{pifont}
\usepackage{amsmath}
\usepackage{amsfonts}
\usepackage{listings}
\usepackage{color}
\usepackage{tabularx}
\usepackage[normalem]{ulem}
\usepackage{caption}
\usepackage{subcaption}
\usepackage{url}
\usepackage[english]{babel}

\graphicspath{{figure/}{figures/}}
\DeclareListParser{\doslashlist}{/}
\newcommand{\mypara}[1]{\smallskip\noindent{\bf {#1}:}~}%
\newcounter{ndnNameComponentCounter}%
\newcommand{\name}[1]{{%
		\setcounter{ndnNameComponentCounter}{0}%
		\renewcommand{\do}[1]{{%
				\ifnumgreater{\value{ndnNameComponentCounter}}{0}{\allowbreak/}{}%
				\ifnumodd{\value{ndnNameComponentCounter}}{}{}%
				\detokenize{##1}}%
			\stepcounter{ndnNameComponentCounter}}%
		``{\fontfamily{cmtt}\small\selectfont\IfBeginWith{#1}{/}{/}{}\doslashlist{#1}}''%
}}
\include{macros}

\setcopyright{none} 

\newcommand{\sysname}{TEB}

\begin{document}
\title{On the Security Bootstrapping in Named Data Networking}
\author{Tianyuan Yu}
\affiliation{%
  \institution{UCLA}
  \city{Los Angeles}
  \country{USA}
}
\email{tianyuan@cs.ucla.edu}

\author{Xinyu Ma}
\affiliation{%
  \institution{UCLA}
  \city{Los Angeles}
  \country{USA}
}
\email{xinyu.ma@cs.ucla.edu}

\author{Hongcheng Xie}
\affiliation{%
  \institution{City University of Hong Kong}
  \city{Hong Kong}
  \country{China}
}
\email{hongcheng.xie@my.cityu.edu.hk}

\author{Xiaohua Jia}
\affiliation{%
  \institution{City University of Hong Kong}
  \city{Hong Kong}
  \country{China}
}
\email{csjia@cityu.edu.hk}

\author{Lixia Zhang}
\affiliation{%
  \institution{UCLA}
  \city{Los Angeles}
  \country{USA}
}
\email{lixia@cs.ucla.edu}
\begin{abstract}
By requiring all data packets been cryptographically authenticatable, the Named Data Networking (NDN) architecture design provides a basic building block for secured networking. 
This basic NDN function requires that all entities in an NDN network go through a security bootstrapping process to obtain the initial security credentials. Recent years have witnessed a number of proposed solutions for NDN security bootstrapping protocols.
Built upon the existing results, in this paper we take the next step to develop a systematic model of security bootstrapping: Trust-domain Entity Bootstrapping (TEB). 
This model is based on the emerging concept of \textit{trust domain} and describes the steps and their dependencies in the bootstrapping process.
We evaluate the expressiveness and sufficiency of this model by using it to describe several current bootstrapping protocols.
\end{abstract}

\maketitle
\pagestyle{plain} 
\input{sections/intro}
\input{sections/background}
\input{sections/bootstrap}
\input{sections/eval}
\input{sections/related}
\input{sections/conclude}

\bibliographystyle{ACM-Reference-Format}
\bibliography{ndn-bootstrap}

\end{document}

%% file: macros.tex
\newboolean{showcomments}
\setboolean{showcomments}{true}
\ifthenelse{\boolean{showcomments}}
{ \newcommand{\mynote}[3]{
    \protect\fbox{\bfseries\sffamily\scriptsize#1}
    {\small$\blacktriangleright$\textsf{\emph{\color{#3}{#2}}}$\blacktriangleleft$}}}
{ \newcommand{\mynote}[3]{}}

\newcommand{\etal}{\textit{et al.}\@\xspace}
\newcommand{\eg}{\textit{e.g.}\@\xspace}
\newcommand{\ie}{\textit{i.e.}\@\xspace}

\definecolor{verde}{rgb}{0.25,0.5,0.35}

\definecolor{verylightgray}{gray}{0.8}



%% file: sections/intro.tex
\section{Introduction}
Named Data Networking (NDN)~\cite{jacobson2009content, zhang2014ndn} architecture provides semantically named and signed data to enable secured data communications.
However, the architecture design needs useful tools to realize the security design.
There are two fundamental requirements to fulfill NDN security design.
First, NDN entities\footnote{NDN entities are applications and all other network communication participants in an NDN network~\cite{zhang2018security}} require \textit{Security Bootstrapping} to obtain necessary security components.
Then, NDN entities also require \textit{Security Support} to manage trust relations and certificates.
Specifically, NDN entities need to know the appropriate certificate to sign data, and the appropriate verification chain to validate signed data.
For example, a data producer may have multiple signing certificates, with different certificates used to sign data under different name prefixes, as indicated by the trust schema~\cite{yu2015schematizing}.
This demands that the Security Support not merely store cryptographic tools, but rather manage cryptographic tools following trust policies.

Recently, a number of security bootstrapping solutions~\cite{nichols-icn2021, li2019ssp, ramani2020ndnviber} have been proposed.
As the understanding of security bootstrapping evolves, in this work we take the next step to extract commonalities from different designs.
By doing so, we intend to obtain a generic model describing the procedures of bootstrapping to understand NDN security better and support software development.
In this work, we make the following contributions.
First, we clarify the concepts of \textit{NDN trust domain}~\cite{nichols-icn2021} and the necessary steps to set up an NDN trust domain.
Second, we develop a systematic understanding of security bootstrapping within a trust domain and propose a generic function model, \textit{Trust-domain Entity Bootstrapping (\sysname)}, to describe the procedures.
Our \sysname\ model is general enough to effectively model the existing protocols.

In the rest of this paper, Section~\ref{sec:background}, revisits the NDN network model, articulates the concept of NDN trust domains, and reviews related works.
We then describe the \sysname\ model and its individual procedures (Section~\ref{sec:bootstrap}), our evaluation based on protocol analysis (Section~\ref{sec:eval}), and lessons learned from the evaluation (Section~\ref{sec:discuss}).
Finally, we summarize our contribution, and mention the remaining questions and future work in Section~\ref{sec:conclude}.


%% file: sections/background.tex
\section{Background and Related Work}
\label{sec:background}
This section briefly reviews the NDN networking model and related work to lay the groundwork for introducing the Trust-domain Entity Bootstrapping TED design.

\subsection{NDN Networking Model}
An NDN is made of connected named entities, with various trust relations among them. 
Entities utilize all available connectivities to exchange named and secured data, and the defined trust schema to authenticate all received data. 
Since the trust schema expresses security policies by defining the relations between the names of data and the names of crypto keys used to sign and encrypt data, an NDN entity $E$ must have a semantically meaningful name to enable schematized trust relations~\cite{yu2015schematizing}.
\begin{figure}[h]
    \centering
	\includegraphics[width=0.47\textwidth]{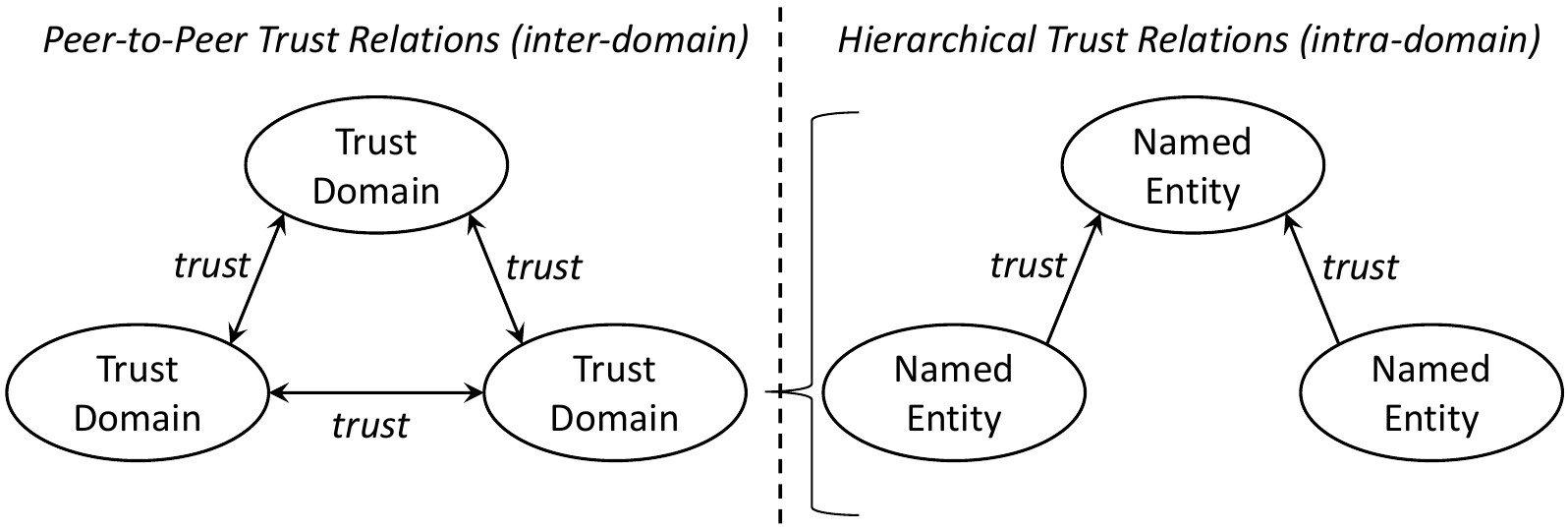}
	\caption{Examples of Trust Relations Among Named Entities}
	\label{fig:trust-relation}
\end{figure}

\begin{figure*}[t]
    \centering
	\includegraphics[width=1.0\textwidth]{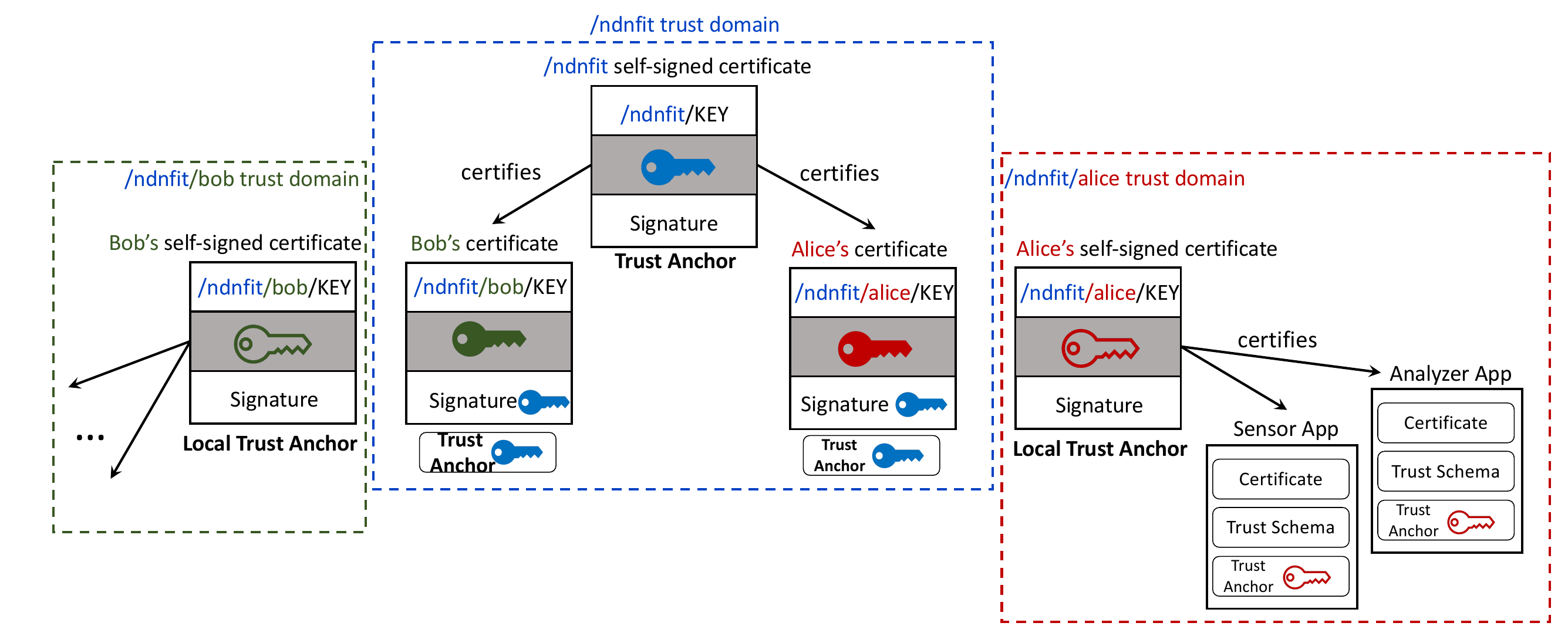}
	\caption[Trust Domain Fig]{The relationships among the trust domains /ndnfit, /ndnfit/alice, and /ndnfit/bob\footnotemark}
	\label{fig:trust-domain}
\end{figure*}

Trust relations can take different models.
Two typical ones are the hierarchical trust relations and the peer-to-peer trust relations, as illustrated in Figure~\ref{fig:trust-relation}.
Named entities that belong to the same administrative domain generally follow a hierarchical trust model~\cite{zhang2020certificate}, where all the entities in the domain share one single trust root.
Independent entities, on the other hand, may take a Web-of-Trust model to establish their trust relations~\cite{yu2014endorsement, gawande2019decentralized}, where they endorse each other's crypto keys in a peer-to-peer fashion.
These two models complement, rather than conflict, each other, and may coexist.
For example, a smart home made of networked IoT devices can use the hierarchical trust model for all IoT devices in a user's home, while neighboring homes may establish peer-to-peer trust relations, if they desire to communicate with each other (e.g. when home-A's ISP has an outage, neighbor home-B may instructs its WiFi router to help forward traffic for home-A, and prohibit all its other devices from interacting with home-A's traffic.

Sine the trust relations in a hierarchical trust model end at one single trust root, realizing a hierarchical trust model requires that each entity $E$ under the same administrative control establishes a \textit{trust anchor} that cryptographically identifies the trusted root, and installs the trust anchor into $E$.
In order to produce authenticatable data inside its domain, $E$ must also have its name(s) certified.
The certified name(s) uniquely identify $E$ in the system and each name's authentication chain terminates at the same trust anchor.
The trust schema, defined by the trust root, limits the signing power of each certificate to a specific data namespace, enabling applications to enforce 
finer-grained security policies for authentication, authorization, and access control.

\subsection{Related Works}\label{sec:related}
Our \sysname\ model is built upon, and further extend, a few pieces of previous work in NDN bootstrapping.
Zhang \etal ~\cite{zhang2018security, zhang2018hostmodel} identified the necessary security components that must be obtained from NDN bootstrapping process.
Later, DCT~\cite{nichols2021trust} defines an NDN \emph{trust domain} as a zero-trust network governed by a single trust anchor and trust schema.
The concept of trust domains helps precisely define the scope of security bootstrapping, i.e. configuring an NDN entity into an NDN trust domain.
Following~\cite{nichols-icn2021}, Yu \etal ~\cite{yu2021enabling} further introduced the concept of a \emph{trust domain controller} as a trust domain's governing entity, and articulated the steps of the security bootstrapping process for three different networking scenarios, where the steps of authentication and naming vary based on the application scenarios.

In this paper, we adopt the concept of trust domain controller and introduce the new concept of \textit{elemental entities}.
We show that \sysname\ as a generalized bootstrapping model can cover all three different networking scenarios described in \cite{yu2021enabling}.

%% file: sections/bootstrap.tex
\section{\sysname\ Model}
\label{sec:bootstrap}
In this section, we formally define the concept of trust domains, and then model the security bootstrapping, by first describing the model in an overview of \sysname\, then introducing each procedure in the model.

\subsection{Trust Domain}
\label{sec:back-trustzone}
The introduction of the NDN \textit{trust domain} concept by~\cite{nichols-icn2021} simplifies the description of trust relation organization.
A \textit{trust domain} is made of a collection of \textit{authorized named entities} under the same administrator's control.
The \textit{trust schema} for an entity $E$ is the set of rules that defines $E$'s trust relations with the others in the same domain.
The entity who controls the trust relations of the domain is the trust domain \textit{controller}.
Specifically, a domain controller can control the security within its domain by 
(i) authenticating and authorizing each new entity $E_{new}$ as a domain member; 
(ii) installing the trust anchor $T$ and the trust schema into $E_{new}$;
(iii) naming $E_{new}$;
(iv) issuing a certificate to $E_{new}$.
We refer to this set of operations as \textit{security bootstrapping}.
To set up an NDN trust domain, one needs to first decide the name of the trust domain, and set up a domain controller which will generate a self-signed certificate under that domain name; this certificate is the domain's trust anchor $T$.
Second, one needs to design the trust domain namespace, and define the trust schema of the domain and individual entities.
Finally, one bootstraps entities into the domain as they become available.
Among all bootstrapped entities, there may exist \textit{elemental entities} to whom the domain controller partially delegates the control function (\eg certificate issuance).
In this case, the domain controller coordinates the elemental entities to manage the security within its domain.
Since this work focuses on intra-domain bootstrapping, below we discuss how 
the domain controller bootstraps $E_{new}$.

In Figure~\ref{fig:trust-domain} (which is adopted from \cite{zhang2018security}), we illustrate the trust domain relationships using a prototype application NDNFit described in~\cite{zhang2016sharing}, which tracks and shares personal fitness activities.
The NDNFit developers start the trust domain \name{/ndnfit} by setting up its trust domain controller, which authenticates and authorizes Alice and Bob as its trust domain members.
It installs the \name{/ndnfit} trust anchor and the trust schema into Alice's and Bob's NDNFit application instances, then issues them the certificates \name{/ndnfit/alice/KEY/123/controller/v=1} and \name{/ndnfit/bob/KEY/223/controller/v=1}.

Alice can set up her own trust domain \name{/ndnfit/alice}, by generating a self-signs \name{/ndnfit/alice} certificate as her domain's trust anchor.
She runs a ''Sensor'' app on her smartphone to collect daily time-location information, and runs an ''Analyzer'' app on her laptop to produce analytics and visualizations from the data produced by ''Sensor''.
The trust domain controller of \name{/ndnfit/alice} installs into ``Sensor'' and ``Analyzer'' their trust anchor, certificates, and trust schema.
After security bootstrapping, the \textit{intra-domain data communications} between the ``Sensor'' app and the ``Analyzer'' app can be secured.

On the left side of the figure, Bob bootstraps his applications in the same way by creating his own trust domain \name{/ndnfit/bob}.
If Alice wants to share her sensor data with Bob, she can enable \textit{inter-domain data communications} by defining a proper trust schema for authentication between two trust domains and data access control.

\subsection{TEB Overview}
\begin{figure}[ht]
    \centering
	\includegraphics[width=0.43\textwidth]{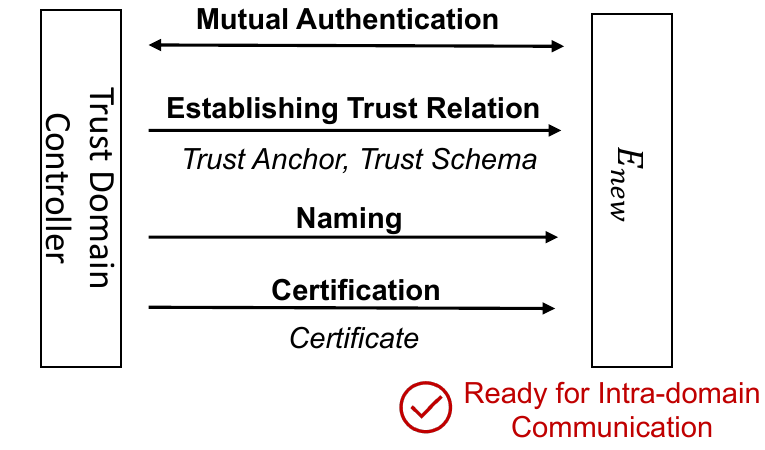}
	\caption{An Overview of the Security Bootstrapping}
	\label{fig:bootstrap-overview}
\end{figure}

\sysname\ is designed as a security bootstrapping model for entities in a trust domain.
\sysname\ consists of four steps (Figure~\ref{fig:bootstrap-overview}).
First, the entity to be bootstrapped, $E_{new}$, and the trust domain controller need to perform mutual authentication in order to securely communicate with each other.
Second, $E_{new}$ needs to establish the trust relation with the trust domain by installing the trust anchor and the trust schema obtained from the controller.
Third, the trust domain controller assigns $E_{new}$ a semantic meaningful NDN name.
Lastly, the trust domain controller issues $E_{new}$ a certificate under its assigned name.
After the above four steps are finished, $E_{new}$ is ready for secure intra-domain data communications. 


\subsection{Controller Authentication}
\label{sec:bootstrap-contauth}
From $E_{new}$'s perspective, the first step of security bootstrapping is to authenticate the trust domain controller and accept it as its authority.
We call this procedure \textit{Controller Authentication}.
Because \textit{authentication relies on pre-established trust relations}, $E_{new}$ and controller need some prior existing trust relation.
We call this trust relation as \textit{Controller Authentication Context} ($CAC$) on $E_{new}$ side, and \textit{$E_{new}$ Authentication Context} ($EAC$) on the controller side.
We further define each party's unique identifier in bootstrapping process as \textit{Controller Identifier} ($ContID$) and \textit{$E_{new}$ Identifier} ($E_{new}ID$), respectively.
As a result of controller authentication, $E_{new}$ generates an $E_{new}Approval$ to represent an entity identified by $ContID$ as its trust domain controller.

We can define the binding between $E_{new}Approval$ and $ContID$ as \textit{Proof of Authority} ($POA$) and the controller authentication procedure as $ContAuth$:
\begin{align*}
    POA=(ContID, E_{new}Approval) \\
    ContAuth: CAC \rightarrow POA
\end{align*}

\mypara{Authentication Context}
Both $E_{new}$ and the trust domain controller need the initial out-of-band trust relation with the other.
For example, $E_{new}$ can securely obtain $CAC$ out-of-band from either the installed application (which embeds some configured trust) or human input at runtime (\eg pre-shared keys or passcode).
In today's TCP/IP network practice, Certificate Authorities (CA) are implicitly authenticated by end-users' trust on the OS and browser vendors, as well as the correct operation of their software.
Whereas, IoT device manufacturers burn the initial key materials into devices as factory settings for $CAC$, to be used for authentication by smart home controllers.
An NDN trust domain can also make use of these existing trust relations to authenticate its trust domain controller to $E_{new}$.
For example, the trust domain controller can host its trust anchor and trust schema at an HTTPS-enabled website and share the URL with $E_{new}$ out-of-band.
In this case, the pre-shared URL, website X.509 certificate~\cite{RFC5280}, and the corresponding CA root certificate together serve as $CAC$.
We discuss this approach in detail in Section~\ref{sec:eval}.

\subsection{Entity Authentication}
\label{sec:bootstrap-enewauth}
In order to join the trust domain, $E_{new}$ can be authenticated and approved by the trust domain controller through the following two steps.
First, the controller verifies $E_{new}$'s trust domain membership.
It checks whether the authentication factor in $EAC$ is acceptable to the current trust domain.
Then, the controller verifies $E_{new}$'s identity with the authentication factor.
After this two-step process, the controller approves $E_{new}$'s to be a trust domain member with the temporary identifier $E_{new}ID$.
We refer to this binding between $E_{new}ID$ and $ContApproval$ as \textit{Proof of Membership} ($POM$).
We can define $POM$ and the $E_{new}$ authentication procedure $E_{new}Auth$ as:
\begin{align*}
    &POM=(E_{new}ID, ContApproval) \\
    &E_{new}Auth: EAC \rightarrow POM
\end{align*}

\mypara{Authentication Context}
The controller must securely obtain $EAC$ before the security bootstrapping.
Similar to $CAC$, obtaining $EAC$ also relies on some trusted existing authentication system, e.g. it can be obtained from either the installed application or from human input at runtime.
Today's applications over the existing TCP/IP architecture take similar approaches to achieve end-user authentication.
For example, today's practice for registering new accounts to websites typically uses email authentication, which derives the user authenticity from the existing (and trusted) email systems.

\subsection{Entity Trust Relations}
\label{sec:bootstrap-trust}
After executing $ContAuth$, $E_{new}$ can use $POA$ to establish the initial trust relation via obtaining and installing the trust anchor and trust schema from the controller.
Obtaining the trust anchor and trust schema enables $E_{new}$ to validate Data packets received within the trust domain, including the certificate issued to it later. 
We name this procedure $E_{new}Trust$:
\begin{align*}
    E_{new}Trust: POA \rightarrow (Trust Anchor, Trust Schema)
\end{align*}

\mypara{Initial Trust Schema}\label{sec:initial-trust-schema}
Since the trust schema may need to be updated over time, $E_{new}Trust$ only installs \textit{initial trust schema} into $E_{new}$ and considers deploying application-specific trust schemas as a post-bootstrapping task for the controller.
The initial trust schema includes all necessary rules to validate $E_{new}$ certificate and future trust schema updates.
As designed in~\cite{zhang2022sovereign}, the trust domain controller can distribute application-specific trust schemas as Data packets after $E_{new}$ bootstrapping.

The initial trust schema can be \textit{implicit}, which means $E_{new}$ by default trust every data produced by the controller until receiving a later trust schema that \textit{explicitly} specifies the data signing relationships.
We discuss the usage of implicit trust schema with details in Section~\ref{sec:eval}. 

\subsection{Entity Naming}
After $E_{new}$'s authentication, the controller assigns $E_{new}$ a name under the trust domain's namespace.
In the $E_{new}$ naming procedure, the controller uses the trust domain's \textit{Naming Convention} ($NameConv$)~\cite{yu2014ndn} to determine $E_{new}$'s name.
The naming convention defines a set of naming rules to facilitate data publication and retrieval.
The controller formally approves the binding of the assigned \textit{Name} to $E_{new}ID$.\footnote{The format of this approval is defined by specific bootstrapping implementations.
}
This binding indicates $E_{new}$'s legitimately possessing $Name$.
We denote this binding as \textit{Proof of Possession} ($POP$),
and define $E_{new}$ naming procedure as:
\begin{align*}
    &POP=(Name, E_{new}ID, ContApproval) \\
    &E_{new}Naming: (POM, NameConv) \rightarrow POP
\end{align*}

At the end of this section, we briefly explain the reason for defining $POP$, rather than the certificate as the $E_{new}Naming$ output.
We also discuss it with more details in Section~\ref{sec:dataflow}.

\mypara{Naming Convention}
An entity name must be \textit{unique} and \textit{semantically meaningful}.
Naming convention remains an active research topic, so far we have identified two commonly observed cases, and one can choose based on what identifiers are used in the authentication step.

The first case is that the controller converts a semantically meaningful and authenticated identifier to an NDN name.
In general, we may be able to satisfy both requirements by making use of the already existing identifiers used in authentication.
In some scenarios, the new entity's authentication identifier (\eg an email address, or a DNS name) semantically encodes its existing trust relations.
For example, Alice is identified by her email address \name{alice@example.com}, which consists of a semantically meaningful domain name and a user name.
The trust domain controller of \name{/ndnfit} can inteprete this email address with the structure \name{user@sld.tld}.
If the trust domain defines its naming convention as \name{/ndnfit/<tld>/<sld>/<user>}\footnote{<> indicates a wildcard name component.},
then the controller can assign Alice the name \name{/ndnfit/com/example/alice}, with the confidence that this name should be unique because email addresses are globally unique.

Another case is to manually assign names.
For example, if Alice is authenticated by her SSH~\cite{ylonen2006secure} public key without meaningful semantics\footnote{A SSH key pair is uniquely identified by its public key (or fingerprint). The default naming convention \name{user@hostname} cannot uniquely identify a key.},
The controller may request human input (such as via out-of-band operations)
to fill in a name.

\mypara{Name Possession}
As stated in the $POP$ definition, $POP$ only indicates the controller's approval on binding a specific name to an $E_{new}ID$, thereby being different from a certificate.
Note that $E_{new}ID$ is not a cryptographic identifier for $E_{new}$,
but rather an identifier exclusively used during bootstrapping before $E_{new}$ can be uniquely identified by its certified name.
$POP$ conceptually decouples the naming procedure from certification, so that the domain certificate issuer can be agnostic to the entity naming convention and takes $POP$ as input to certify $E_{new}$.

\subsection{Certification}
After $E_{new}$ obtains its name assignment from $E_{new}Naming$, it needs to obtain the certificate from $POP$.
The certification procedure requires $POP$ for name input, and \textit{Certification Context} ($CertC$) as certificate issuer context.
Moreover, the procedure needs $Trust Anchor$ and $Trust Schema$ to validate the issued certificate conforming to the trust schema of the domain.
In our model, we describe certification as the $E_{new}Cert$ procedure:
\begin{align*}
    E_{new}Cert: &(POP, CertC, Trust Anchor, Trust Schema) \\
    &\rightarrow Certificate
\end{align*}

\mypara{Certificate Issuer}
Inside a trust domain, either the domain controller itself, or another delegated entity, needs to be responsible for certificate issuance and revocation, as well as making them available by publishing to some repository~\cite{ndnpythonrepo}.
That is, a certificate issuer is not necessarily the trust domain controller itself.
During certificate issuance, the issuer validates $POP$ and binds the contained name assignment to $E_{new}$'s public key.
Today's CAs, such as Let's Encrypt~\cite{aas2019let}, issue domain-validated certificates (DV) following a similar logic.
The domain validation process validates the requester's proof of DNS domain possession (\eg publishing a DNS TXT Record) by trusting the global routing system for delivering all validation requests to correct destinations.
NDN certificate issuers rely on $POP$ to provide authenticated name--entity binding.
Later in Section~\ref{sec:eval}, we show individual security bootstrapping protocols have their own $POP$ realizations, such as session encryption keys and temporary certificates. 

\subsection{\sysname\ Dataflow Graph}
\label{sec:dataflow}
The previous subsections explained individual procedures in security bootstrapping.
Now we show how the procedures of \sysname\ work together to form a framework for the security bootstrapping process.

\begin{figure}[ht]
    \centering
	\includegraphics[width=0.43\textwidth]{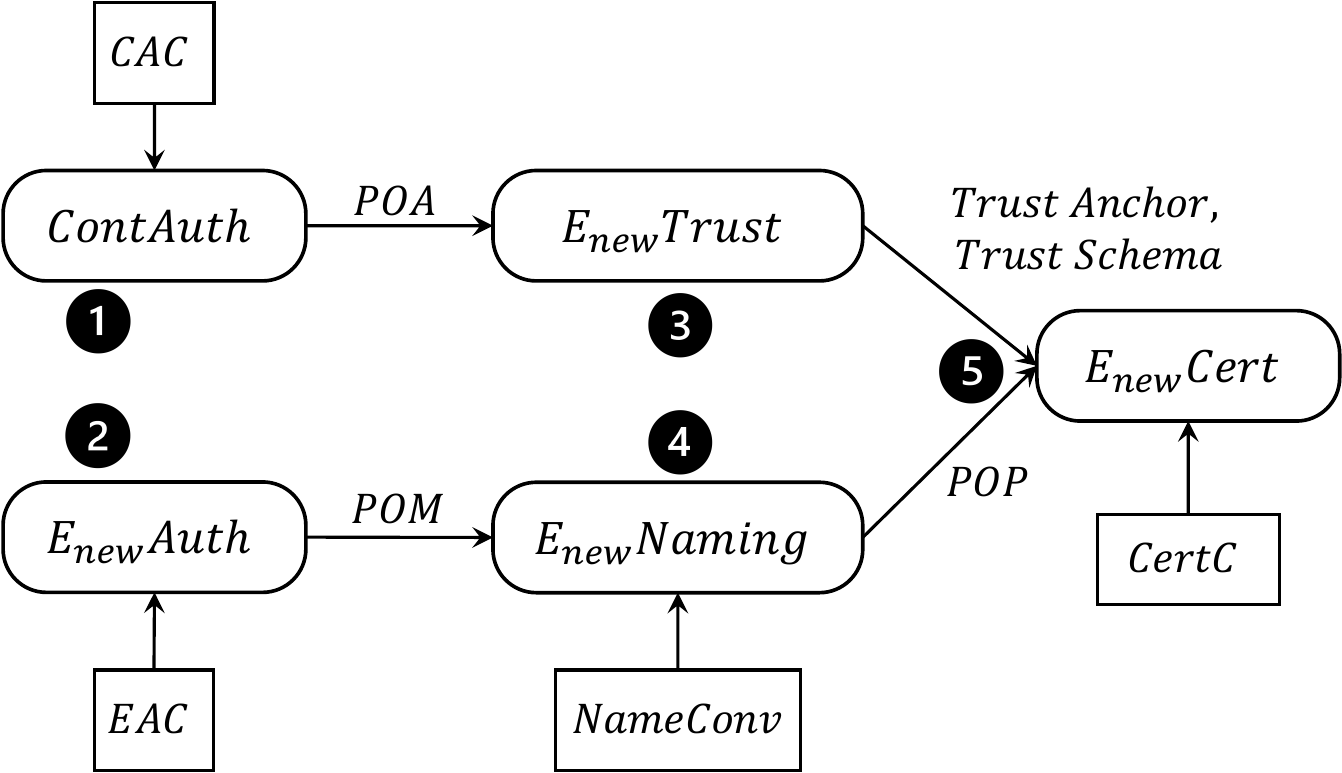}
	\caption{Dataflow Graph of the Bootstrapping Model}
	\label{fig:bootstrap-dataflow}
\end{figure}

Figure~\ref{fig:bootstrap-dataflow} shows the dataflow graph of security bootstrapping, where $CAC$ and $EAC$ represent the pre-existing trust relation between the domain controller and $E_{new}$.
When the security bootstrapping starts, $ContAuth$ \ding{202} and $E_{new}Auth$ \ding{203} generate $POA$ and $POM$, respectively.
Then, $E_{new}Trust$ \ding{204} obtains $Trust Anchor$ and $Trust Schema$.
In parrallel, $E_{new}Naming$ \ding{205} takes $POM$ and $NameConv$ as input to produce $POP$.
As the final step, $E_{new}Cert$ \ding{206} utilizes $POP$ and $CertC$ to obtain a certificate, and validates it with $Trust Anchor$ and $Trust Schema$.


\mypara{Modularized Bootstrapping Design}
Since the dataflow graph reveals the procedure dependencies in bootstrapping, it informs a new possibility in the bootstrapping protocol design.
That is, the two-way authentication can be decoupled, so that one can authenticate and name $E_{new}$ before $E_{new}$ accepts the trust anchor and the initial trust schema.
The certification can be simplified to be a procedure that binds $E_{new}$ name assignment to a public-private key pair, which requires accomplishing authentication and naming a prior, rather than in real-time.

The dataflow graph also enables modularity in the protocol design and implementation.
Because a valid security bootstrapping process is a set of procedures that follow the dataflow graph, developers can realize each procedure separately with any execution order that satisfies the \sysname\ dataflow graph.
Then a \sysname\ implementation bootstraps $E_{new}$ by executing individual procedures in the specified order.
One can also summarize commonly used procedure implementations and plug them into various \sysname-based bootstrapping designs.
For example, an $E_{new}Auth$ implementation that authenticates $E_{new}$ based on email addresses can be shared among multiple trust domain bootstrapping solutions to ease the protocol development.

%% file: sections/eval.tex
\section{Evaluation}
\label{sec:eval}
We evaluate our model by analyzing several security bootstrapping solutions.
\begin{table*}[ht]
    \centering
    \newcolumntype{C}{>{\raggedright\let\newline\\\arraybackslash\hspace{0pt}}m{0.25\linewidth} }
	\newcolumntype{D}{>{\raggedright\arraybackslash} m{0.45\linewidth} }
    \newcolumntype{E}{>{\raggedright\arraybackslash} m{0.45\linewidth} }
    \resizebox{\textwidth}{!}{
        \begin{tabular}{CDE}
            \hline
            \textbf{Operations} & \textbf{SSP Devices (D)} & \textbf{SSP Controller (C)} \\ \hline \hline
            Out-of-band (OOB) & Obtaining $CAC$ & Obtaining $EAC$, $NameConv$ \\ \hline \hline
            D $\Rightarrow$ C: sign-on Interest & $E_{new}ID$ Generation & $E_{new}Auth(EAC) \rightarrow POM$ \\ \hline
            D $\Leftarrow$ C: sign-on Data & $E_{new}Trust(ContAuth(CAC)) \rightarrow Trust Anchor, Trust Schema$, $CertC$ Generation & $ContID$ Generation\\ \hline
            D $\Rightarrow$ C: certificate request Interest & & $E_{new}Naming(POM, NameConv) \rightarrow POP$ \\ \hline
            D $\Leftarrow$ C: certificate Data & $E_{new}Cert(POP, CertC, Trust Anchor, Trust Schema) \rightarrow Certificate$ & \\ \hline
        \end{tabular}
    }
	\caption{SSP Security Bootstrapping Model}
	\label{table:ssp-model}
\end{table*}
\begin{table*}[ht]
    \centering
    \newcolumntype{C}{>{\raggedright\let\newline\\\arraybackslash\hspace{0pt}}m{0.25\linewidth} }
	\newcolumntype{D}{>{\raggedright\arraybackslash} m{0.45\linewidth} }
    \newcolumntype{E}{>{\raggedright\arraybackslash} m{0.45\linewidth} }
    \resizebox{\textwidth}{!}{
        \begin{tabular}{CDE}
            \hline
            \textbf{Operations} & \textbf{NDN Testbed Users (U)} & \textbf{NDN Testbed CA (C)} \\ \hline \hline
            OOB & $Trust Anchor, Trust Schema$ installation & Obtaining $EAC$, $NameConv$ \\ \hline \hline
            U $\Leftrightarrow$ C: NEW Interest-Data & Obtaning $E_{new}ID$ & $E_{new}ID$ Generation \\ \hline
            U $\Leftrightarrow$ C: CHALLENGE Interest-Data (Round-trip 1) & Obtaning $POM$ & $E_{new}Auth(EAC) \rightarrow POM$ \\ \hline
            U $\Leftrightarrow$ C: CHALLENGE Interest-Data (Round-trip 2) & Obtaning $POP$, $CertC$ & $E_{new}Naming(POM, NameConv) \rightarrow POP$ \\ \hline
            U $\Leftrightarrow$ C: Certificate Interest-Data & $E_{new}Cert(POP, CertC, Trust Anchor, Trust Schema) \rightarrow Certificate$ & \\ \hline
        \end{tabular}
    }
	\caption{NDN Testbed Security Bootstrapping Model}
	\label{table:testbed-model}
\end{table*}

\mypara{SSP}\label{sec:eval-ssp}
SSP~\cite{li2019ssp} is a security bootstrapping protocol aiming at smart home device.
In SSP, smart home devices (\ie $E_{new}$) pre-share a QR code to the home controller (\ie trust domain controller).
The QR code includes a public key, a symmetric key and the device identifier.
In this case, $EAC$ is the shared public key, symmetric key and device identifier,
and $CAC$ is the corresponding private key and symmetric key.

Firstly, the device initiates security bootstrapping by broadcasting a sign-on Interest packet (\ding{202}).
The sign-on Interest carries the device identifier, device capability, and a nonce $N1$, as $E_{new}ID$, signed with the device's private key.
Before broadcasting the Interest, the device hashes the all the four parameters as a whole and appends it as the last name component.
After receiving the sign-on Interest, the controller verifies the signature with the pre-shared public key (\ding{203}).
Upon successful verification, the controller replies an sign-on Data packet with the same name.
The sign-on Data encapsulates $Trust Anchor$ (controller's self-signed certificate), $Trust Schema$ (implicit), and nonce $N2$ as $ContID$, signed with the shared symmetric key.
Because the sign-on Data name carries the same digest of parameters with the sign-on Interest, the data signature ensures the device membership in the smart home trust domain.
Meanwhile the symmetrically signed data signature is verifiable to the device, thereby the sign-on Data packet is both $POM$ and $POA$.

Secondly, the device receives and validates the sign-on Data.
It installs $Trust Anchor$ and $Trust Schema$ from the sign-on Data content (\ding{204}) and performs ECDH between its private key with the controller's public key.
The temporary symmetric key obtained from ECDH is the $CertC$ used later.

Thirdly, the device broadcasts a certificate request Interest (\ding{205}).
The certificate request Interest carries the device identifier, $Trust Anchor$ digest, $N1$, $N2$, and a signature from the device.
Similar to the sign-on Interest, the device appends the digest of parameters to the name.
Upon successful verification, the controller assigns a name to the device according to $NameConv$, generates a new key pair for the device, and certifies the device's new public key under the name assignment.
It also performs ECDH between the controller private key and the device pre-shared public key to negotiate a temporary symmetric key for private key encryption.
Then the controller replies with a certificate Data carrying the same name.
The certificate Data contains the device's certificate and the encrypted device's new private key and is signed by the shared symmetric key.
As $POP$, its data signature authenticates the binding between $N1$ and the device name assignment.

Finally, the device validates the received certificate Data, decrypts the new private key with $CertC$, and installs the certificate (\ding{206}) to complete the security bootstrapping process.
We summarize the SSP bootstrapping model in Table~\ref{table:ssp-model}.

\mypara{NDN Testbed}\label{sec:eval-testbed}
NDN Testbed distributes the trust anchor through NDN Website~\cite{ndn-testbed} and NDNCERT codebase~\cite{ndncert-code} that implements the certificate issuance protocol of Testbed.
Testbed users authenticate, obtain and install $Trust Anchor$ and $Trust Schema$ out-of-band (\ding{202} \ding{204}), then follows the NDNCERT~\cite{zhang2020certificate, ndncert-spec} protocol to request a certificate on Testbed.
NDN Testbed bootstraps users via authenticating users' email addresses and issuing certificates to users.
In the bootstrapping process, a user runs the NDNCERT protocol and acts as a certificate requester.

Firstly, the certificate requester takes $EAC$ to initiate user authentication (\ding{203}).
$EAC$ contains the user's email address and the NDN Testbed Certificate Authority (CA) prefix.
The requester sends a NEW Interest that carries an ECDH public key, the public key to be certified, and a signature generated by the corresponding private key.
In reply, the CA sends a Data packet carrying $RequestID$, as $E_{new}ID$, which is randomly picked to identify the certificate request instance 
Note that both Interest and Data contain the public information used for the ECDH key agreement and are signed with the timestamp and nonce to prevent replay attack.

Secondly, the requester sends a CHALLENGE Interest 1 carrying $RequestID$ and the user's email address which is encrypted using the key negotiated from the NEW Interest-Data exchange.
The CA obtains the email address from the CHALLENGE Interest 1 and decides whether it accepts this email address in Testbed.
Afterward, it sends a randomly generated PIN to the email address and replies CHALLENGE Data 1.
Similar to the NEW step, CHALLENGE Interest and Data packets are signed by the sender and verified by the receiver.
As $POM$, the CHALLENGE Data 1 binds the $RequestID$ with the CA signature.
The Testbed user obtains the PIN out-of-band and inputs it to the requester.

Thirdly, the requester sends CHALLENGE Interest 2 carrying $RequestID$ and the encrypted PIN. 
Upon successful PIN verification, the CA assigns a name, certifies it under the requester public key, and replies CHALLENGE Data 2 (\ding{205}).
As $POP$, CHALLENGE Data 2 contains $RequestID$, the encrypted certificate name, and the CA signature.
Thereafter, the requester obtains and decrypts the $CertC$ from $POP$.
$CertC$ consists of a certificate name and a forwarding hint to facilitate certificate retrieval.

As the final step, the requester expresses Interest with the certificate name and a forwarding hint to retrieve the issued certificate to be installed locally (\ding{206}).
We summarize the NDN Testbed bootstrapping model in Table~\ref{table:testbed-model}.

\begin{table*}[ht]
    \centering
    \newcolumntype{C}{>{\raggedright\let\newline\\\arraybackslash\hspace{0pt}}m{0.25\linewidth} }
	\newcolumntype{D}{>{\raggedright\arraybackslash} m{0.45\linewidth} }
    \newcolumntype{E}{>{\raggedright\arraybackslash} m{0.45\linewidth} }
    \resizebox{\textwidth}{!}{
        \begin{tabular}{CDE}
            \hline
            \textbf{Operations} & \textbf{NDNViber Devices (D)} & \textbf{NDNViber Controller (C)} \\ \hline \hline
            OOB & Obtaining $EAC$, $CAC$ & Obtaining $NameConv$ \\ \hline \hline
            (Vibration Channel) D $\Leftrightarrow$ C: TRIGGER Interest-Data & $ContAuth(CAC) \rightarrow POA$ & Obtaining $EAC$ \\ \hline
            D $\Leftrightarrow$ C: ANCHOR Interest-Data & $E_{new}Trust(ContAuth(CAC)) \rightarrow Trust Anchor, Trust Schema$ & \\ \hline
            D $\Leftrightarrow$ C: NDNCERT Interest-Data (multiple rounds) & Obtaining $POM$, $POP$, $CertC$ & $E_{new}Auth(EAC) \rightarrow POM$, $E_{new}Naming(POM, NameConv)\rightarrow POP$ \\ \hline
            D $\Leftrightarrow$ C: Certificate Interest-Data & $E_{new}Cert(POP, CertC, Trust Anchor, Trust Schema) \rightarrow Certificate$ & \\ \hline
        \end{tabular}
    }
	\caption{NDNViber Security Bootstrapping Model}
	\label{table:ndnviber-model}
\end{table*}

\begin{table*}[ht]
    \centering
    \newcolumntype{C}{>{\raggedright\let\newline\\\arraybackslash\hspace{0pt}}m{0.25\linewidth} }
	\newcolumntype{D}{>{\raggedright\arraybackslash} m{0.4\linewidth} }
    \newcolumntype{E}{>{\raggedright\arraybackslash} m{0.25\linewidth} }
    \resizebox{\textwidth}{!}{
        \begin{tabular}{CDDE}
            \hline
            \textbf{Operations} & \textbf{PION Devices (D)} & \textbf{PION Authenticator (A)} & \textbf{Controller (C)} \\ \hline \hline
            OOB & Obtaining $CAC$ & Obtaining $EAC$, $NameConv$ & \\ \hline \hline
            A $\Leftrightarrow$ D: PAKE Interest-Data  & & $E_{new}Auth(EAC) \rightarrow POM$ & \\ \hline
            A $\Leftrightarrow$ D: CONFIRM Interest-Data  & $ContAuth(CAC) \rightarrow POA$, $E_{new}Trust(POA) \rightarrow TrustAnchor, TrustSchema$, Obtaining $POM$, $CertC$ & \\ \hline
            A $\Leftrightarrow$ D: CREDENTIAL Interest-Data  & & $E_{new}Naming(POM, NameConv) \rightarrow POP$ & \\ \hline
            A $\Leftrightarrow$ D: Certificate Interest-Data & Obtaining $POP$ & & \\ \hline
            D $\Leftrightarrow$ C: NDNCERT Interest-Data (multiple rounds) & Obtaining $POP'$, $CertC$ & & $E_{new}Cert_1(POP) \rightarrow POP'$ \\ \hline
            D $\Leftrightarrow$ C: Certificate Interest-Data & $E_{new}Cert_2(POP', CertC, Trust Anchor, Trust Schema) \rightarrow Certificate$ & & \\ \hline
        \end{tabular}
    }
	\caption{PION Security Bootstrapping Model}
	\label{table:pion-model}
\end{table*}

\mypara{NDNViber}\label{sec:eval-viber}
NDNViber\cite{ramani2020ndnviber} is an automated bootstrapping protocol for IoT devices.
It uses the vibration channel in physical proximity to realize the two-way authentication.
The vibration channel information (\eg coding scheme) is the $CAC$ on $E_{new}$ side.
NDNViber controller initiates the bootstrapping process by expressing a TRIGGER Interest in the vibration channel.
The TRIGGER Interest includes the trust domain name and a temporary encryption key.
Because of the secrecy of the vibration channel, the NDNViber device considers the TRIGGER Interest as $POA$ (\ding{202}) and replies with the pre-installed device identifier as $EAC$.
When the controller receives the corresponding TRIGGER Data, it obtains the $EAC$ for device authentication.

After the TRIGGER Interest-Data exchange, the device expresses ANCHOR Interest over the traditional channels and uses the temporary encryption key from $POA$ to obtain $TrustAnchor$ and $TrustSchema$ (implicit) (\ding{204}).
Finally, the device runs the NDNCERT protocol to obtain $POP$ and $Certificate$.
The controller authenticates (\ding{203}) and names the device based on $EAC$ and $NameConv$, and issues certificates (\ding{206}) to the device via the NDNCERT CHALLENGE Interest-Data exchanges.
We omit the detailed analysis because of its similarity to the NDN Testbed $E_{new}Auth$, $E_{new}Naming$ and $E_{new}Cert$.
The main difference is that $E_{new}Auth$ is over the vibration channel.
We model the protocol in Table~\ref{table:ndnviber-model}.

\mypara{PION}\label{sec:eval-pion}
PION~\cite{davide2022pion, pion-spec} is a password-based security bootstrapping for IoT devices.
Different from the aforementioned protocols, in PION, the controller delegates authentication and naming to an elemental entity called the PION authenticator.
The PION authenticator authenticates and names $E_{new}$ with a temporary certificate as $POP$, and $E_{new}$ further uses the temporary certificate to apply the formal certificate from the controller.
It realizes the two-way authentication through the first pre-shared password as both $EAC$ and $CAC$, then derives a shared secret following the SPAKE2~\cite{irtf-cfrg-spake2-26} scheme to secure communications during bootstrapping.

As a trusted elemental entity, the PION authenticator initiates the bootstrapping process by sending PAKE Interest, which includes the authenticator SPAKE2 public share \textit{pA}.
The device receives PAKE Interest and processes \textit{pA} based on the password, then replies PAKE Data which includes its public share \textit{pB} and key confirmation message \textit{cB}.
When the authenticator receives this PAKE Data, the authenticator 
(i) processes \textit{pB}, verfies \textit{cB}, generates its confirmation message \textit{cA};
(ii) assigns \textit{SID} as $E_{new}ID$, and names $E_{new}$ according to the $NameConv$ obtained OOB;
(iii) derives a temporary encryption key \textit{Ke} from the previous SPAKE2 exchanges;
(iv) sends a CONFIRM Interest as $POM$ (\ding{203}) that carries \textit{SID} and \textit{Ke} encrypted $E_{new}$ name assignment.
Upon successful \textit{cA} verfication, the device accepts the received CONFIRM Interest as $POM$ and $POA$ (\ding{202}).
It extracts the $Trust Anchor$ and the $Trust Schema$ (implicit) from the $POA$ (\ding{204}), and replies a CONFIRM Data.
The CONFIRM Data includes \textit{SID} and \textit{Ke} encrypted $E_{new}$ self-signed certificate.
The PION authenticator receives this Interest, signs a temporary $E_{new}$ certificate as $POP$, and sends the \textit{Ke} encrypted certificate name back to the device through a CREDENTIAL Interest (\ding{205}).
In order to obtain $POP$, the device uses \textit{Ke} to decrypt the certificate name from CREDENTIAL Interest with, then fetch and temporary certificate.

Afterwards, the device follows the NDNCERT protocol and uses $POP$ and $CertC$ to obtain the name of the formally issued certificate from the trust domain controller (\ding{206}).
In PION, 
Since the temporary certificate as $POP$ already binds the controller's approval on $E_{new}$'s identifier to the name, we model the CHALLENGE Data of the NDNCERT exchanges in PION as $POP'$.
The device first follows the NDNCERT protocol to obtain $POP'$ and update $CertC$ to $CertC'$, then uses $CertC'$ to decrypt the certificate name from $POP'$ and expresses Interest to fetch the formally issued certificate.
We use the following two subprocedures to describe the $E_{new}Cert$.
\begin{align*}
E_{new}Cert_1: &(POP) \rightarrow POP' \\
E_{new}Cert_2: &(POP', CertC, TrustAnchor, TrustSchema) \\
&\rightarrow Certificate
\end{align*}

Table~\ref{table:pion-model} shows the PION protocol analysis.

\mypara{DCT}\label{sec:eval-dct}
DCT~\cite{nichols-icn2021} is a data-centric toolkit for secure NDN IoT applications.
In order to bootstrap entities inside a DCT trust domain, developers specify the trust anchor and the trust schema for a trust domain, and generate \textit{Identity Bundle} for each entity.
The Identity Bundle contains the trust anchor, trust schema, certificate chain, and the corresponding private keys.
Human operators bootstrap each entity by securely installing its Identity Bundle.
Specifically, DCT suggests operators install bundles via command line within the development environment, and it leaves other installation mechanisms to individual trust domain deployment.

Although DCT bootstrapping does not involve explicit data communications, it still follows the \sysname\ model.
Securely installing bundles requires the trust domain controller and $E_{new}$ mutually authenticating each other (\ding{202} \ding{203}).
Then the parties can establish a secured channel in between and enable the controller securely transfer (\eg command line) the bundle into $E_{new}$ (\ding{204} \ding{205} \ding{206}).

Besides the above representative bootstrapping protocols, \cite{yu2021enabling} proposes three cases of security bootstrapping.
The first case assumes a secure environment between the trust domain controller and the new entity ($E_{new}$), which is similar to the DCT deployment scenario (Section~\ref{sec:eval-dct}).
While the second case assumes the new entity is within the trust domain controller's physical vicinity, where our analyses of SSP (Section~\ref{sec:eval-ssp}) and NDNViber (Section~\ref{sec:eval-viber}) are applicable.
The third case leverages existing authentication solutions in today's Internet and NDNCERT for both authentication and certification.
This case naturally fits into our analysis of the NDN Testbed bootstrapping model (Section~\ref{sec:eval-testbed}).
Therefore, \sysname\ covers all three different networking scenarios in \cite{yu2021enabling}.

%% file: sections/related.tex
\section{Discussion}
\label{sec:discuss}
Having described the \sysname\ design and evaluation, in this section we discuss \sysname's design decisions and the lessons learned from the evaluation.

\subsection{Authentication and Certification}
A certificate represents the trust domain controller's endorsement of the binding between an entity's name and its public key.
In order to issue a certificate, the controller must
(i) authenticate the public key owner and binds it to a name; and
(ii) endorse the name--key binding through digital signature.
If a domain supports multiple certificate issuers, these issuers can share the same $EAC$ and $NameConv$.

\sysname\ breaks the aforementioned two steps into three procedures: $E_{new}Auth$, $E_{new}Naming$, and $E_{new}Cert$.
Since each procedure serves one specific purpose, their realizations are relatively simple.
In this regard, PION separates the name authenticator from the certificate issuer.
The authenticator issues a temporary certificate that carries the authenticated name assignment.
Consequently, the certificate issuer can be $EAC$ and $NameConv$ agnostic by re-signing a formal certificate to the public key owner.

\subsection{Multi-Named Entities}
\sysname\ allows an entity to have multiple names within a trust domain, with each name uniquely identifying the entity.
A typical scenario that may need multiple names for one entity is Name-based Access Control (NAC)~\cite{zhang2018nac}.
In the example of Figure~\ref{fig:trust-domain}, the NDNFit controller can assign Alice two additional names \name{/ndnfit/alice/admin} and \name{/ndnfit/alice/customer}, each representing a different user role.
When the controller specifies access control policies, it can let Alice request decryption keys based on role names.

To support multi-named entities, the controller assigns multiple names in $E_{new}Naming$.
The certificate issuer of each name takes actions according to the trust schema and $E_{new}Cert$ realization.
A possible approach is to certify \name{/ndnfit/alice} first, and then to certify the two additional names.
We observed the same design pattern in DCT, where the DCT controller certifies devices and the devices further sign their capability certificates (e.g., light).
Therefore, each DCT device may have multiple names, depending on the capability it equips with.

\subsection{Trust Schema: Implicit vs. Explict}
As we mentioned earlier, the trust schema is a critical security component that must be obtained during bootstrapping.
Designing the trust schema for a trust domain requires understanding both application semantics and security requirements.

\sysname\ allows the initial trust schema to be implicit (Section~\ref{sec:initial-trust-schema}), which provides a shortcut for the out-of-band configuration of the new entity before bootstrapping.
However, since the implicit trust schema only allows data produced by the controlled to be trusted, the new entity cannot securely communicate with any other entity, until it learns how to validate data produced by others.
In other words, distributing an explicit and finer-tuned trust schema after security bootstrapping is necessary, which leads to extra overheads.

We believe that the best practice is to explicitly specify an initial trust schema in $E_{new}Trust$.
DCT achieves this by putting the certificate chain and the trust schema into the identity bundle.
All entities in a trust domain are able to validate others' data publication once obtaining the bundle.

%% file: sections/conclude.tex
\section{Conclusion and Future Work}
\label{sec:conclude}
NDN's network model requires all named entities to establish trust relations with others.
As explained in~\cite{yu2021enabling}, security bootstrapping fulfills this requirement by answering two basic questions:
from where a new entity can obtain its name(s) and security credentials, and
how the initial trust relations can be configured into the entity.
However, our observations on recent NDN application development suggest that it is non-trivial for developers to understand the necessary procedures to perform security bootstrapping and properly implement entity bootstrapping operations.

To address the above issues, we proposed \sysname\ to formally define the goal and procedures of security bootstrapping based on the concept of NDN trust domain.
We evaluated \sysname\ against various existing security bootstrapping solutions and show that those individual protocols all fit into the \sysname\ model.
Our experience so far gives us confidence in the model's generality in support security bootstrapping for most, if not all, application scenarios in a trust domain.
We believe we have contributed a meaningful step towards a reusable approach in bootstrapping protocol designs:
defining abstract variables, realizing logical procedures, and executing procedures based on functional dependencies.

Note that security bootstrapping is only the first step in securing data communications.
One remaining research question is where to store these security components and how to automatically execute trust policies.
Since \sysname\ focuses on securing intra-domain data communications, another question is how to establish inter-domain trust relations.
Communicating with semantically named and signed data empowers one to leverage naming conventions and automate security workflows.
However, we need well-integrated and easy-to-use implementations that realize such automation, and hide security primitives from the application developers.
As our future work, we plan to develop solutions to the two aforementioned questions, and provide a complete framework to realize the \sysname\ model.

%% file: ndn-bootstrap.bbl

\begin{thebibliography}{27}


\ifx \showCODEN    \undefined \def \showCODEN     #1{\unskip}     \fi
\ifx \showDOI      \undefined \def \showDOI       #1{#1}\fi
\ifx \showISBNx    \undefined \def \showISBNx     #1{\unskip}     \fi
\ifx \showISBNxiii \undefined \def \showISBNxiii  #1{\unskip}     \fi
\ifx \showISSN     \undefined \def \showISSN      #1{\unskip}     \fi
\ifx \showLCCN     \undefined \def \showLCCN      #1{\unskip}     \fi
\ifx \shownote     \undefined \def \shownote      #1{#1}          \fi
\ifx \showarticletitle \undefined \def \showarticletitle #1{#1}   \fi
\ifx \showURL      \undefined \def \showURL       {\relax}        \fi
\providecommand\bibfield[2]{#2}
\providecommand\bibinfo[2]{#2}
\providecommand\natexlab[1]{#1}
\providecommand\showeprint[2][]{arXiv:#2}

\bibitem[Aas et~al\mbox{.}(2019)]%
        {aas2019let}
\bibfield{author}{\bibinfo{person}{Josh Aas}, \bibinfo{person}{Richard Barnes},
  \bibinfo{person}{Benton Case}, \bibinfo{person}{Zakir Durumeric},
  \bibinfo{person}{Peter Eckersley}, \bibinfo{person}{Alan Flores-L{\'o}pez},
  \bibinfo{person}{J~Alex Halderman}, \bibinfo{person}{Jacob Hoffman-Andrews},
  \bibinfo{person}{James Kasten}, \bibinfo{person}{Eric Rescorla},
  {et~al\mbox{.}}} \bibinfo{year}{2019}\natexlab{}.
\newblock \showarticletitle{Let's Encrypt: an automated certificate authority
  to encrypt the entire web}. In \bibinfo{booktitle}{\emph{Proceedings of the
  2019 ACM SIGSAC Conference on Computer and Communications Security}}.
  \bibinfo{pages}{2473--2487}.
\newblock


\bibitem[Cooper et~al\mbox{.}(2008)]%
        {RFC5280}
\bibfield{author}{\bibinfo{person}{D. Cooper}, \bibinfo{person}{S. Santesson},
  \bibinfo{person}{S. Farrell}, \bibinfo{person}{S. Boeyen},
  \bibinfo{person}{R. Housley}, {and} \bibinfo{person}{W. Polk}.}
  \bibinfo{year}{2008}\natexlab{}.
\newblock \bibinfo{booktitle}{\emph{Internet X.509 Public Key Infrastructure
  Certificate and Certificate Revocation List (CRL) Profile}}.
\newblock \bibinfo{type}{RFC} 5280. \bibinfo{institution}{RFC Editor}.
\newblock
\showISSN{2070-1721}


\bibitem[Gawande et~al\mbox{.}(2019)]%
        {gawande2019decentralized}
\bibfield{author}{\bibinfo{person}{Ashlesh Gawande}, \bibinfo{person}{Jeremy
  Clark}, \bibinfo{person}{Damian Coomes}, {and} \bibinfo{person}{Lan Wang}.}
  \bibinfo{year}{2019}\natexlab{}.
\newblock \showarticletitle{Decentralized and secure multimedia sharing
  application over named data networking}. In
  \bibinfo{booktitle}{\emph{Proceedings of the 6th ACM Conference on
  Information-Centric Networking}}. \bibinfo{pages}{19--29}.
\newblock


\bibitem[Jacobson et~al\mbox{.}(2009)]%
        {jacobson2009content}
\bibfield{author}{\bibinfo{person}{Van Jacobson}, \bibinfo{person}{Diana~K.
  Smetters}, \bibinfo{person}{James~D. Thornton}, \bibinfo{person}{Michael~F.
  Plass}, \bibinfo{person}{Nicholas~H. Briggs}, {and}
  \bibinfo{person}{Rebecca~L. Braynard}.} \bibinfo{year}{2009}\natexlab{}.
\newblock \showarticletitle{Networking Named Content}. In
  \bibinfo{booktitle}{\emph{Proceedings of the 5th International Conference on
  Emerging Networking Experiments and Technologies}} (Rome, Italy)
  \emph{(\bibinfo{series}{CoNEXT '09})}. \bibinfo{publisher}{Association for
  Computing Machinery}, \bibinfo{address}{New York, NY, USA},
  \bibinfo{pages}{1--12}.
\newblock
\showISBNx{9781605586366}


\bibitem[Ladd and Kaduk(2022)]%
        {irtf-cfrg-spake2-26}
\bibfield{author}{\bibinfo{person}{Watson Ladd} {and} \bibinfo{person}{Benjamin
  Kaduk}.} \bibinfo{year}{2022}\natexlab{}.
\newblock \bibinfo{booktitle}{\emph{{SPAKE2, a PAKE}}}.
\newblock \bibinfo{type}{Internet-Draft} draft-irtf-cfrg-spake2-26.
  \bibinfo{institution}{Internet Engineering Task Force}.
\newblock
\newblock
\shownote{Work in Progress}.


\bibitem[{Li} et~al\mbox{.}(2019)]%
        {li2019ssp}
\bibfield{author}{\bibinfo{person}{Y. {Li}}, \bibinfo{person}{Z. {Zhang}},
  \bibinfo{person}{X. {Wang}}, \bibinfo{person}{E. {Lu}}, \bibinfo{person}{D.
  {Zhang}}, {and} \bibinfo{person}{L. {Zhang}}.}
  \bibinfo{year}{2019}\natexlab{}.
\newblock \showarticletitle{A Secure Sign-On Protocol for Smart Homes over
  Named Data Networking}.
\newblock \bibinfo{journal}{\emph{IEEE Communications Magazine}}
  \bibinfo{volume}{57}, \bibinfo{number}{7} (\bibinfo{date}{July}
  \bibinfo{year}{2019}), \bibinfo{pages}{62--68}.
\newblock
\showISSN{0163-6804}


\bibitem[{Named Data Networking Project}(2022)]%
        {ndn-testbed}
\bibfield{author}{\bibinfo{person}{{Named Data Networking Project}}.}
  \bibinfo{year}{2022}\natexlab{}.
\newblock \bibinfo{title}{NDN Testbed}.
\newblock
\newblock
\newblock
\shownote{\url{https://named-data.net/ndn-testbed/}}.


\bibitem[Nichols(2021a)]%
        {nichols-icn2021}
\bibfield{author}{\bibinfo{person}{Kathleen Nichols}.}
  \bibinfo{year}{2021}\natexlab{a}.
\newblock \showarticletitle{Trust Schemas and ICN: Key to Secure Home IoT}
  \emph{(\bibinfo{series}{ICN '21})}. \bibinfo{publisher}{Association for
  Computing Machinery}, \bibinfo{address}{New York, NY, USA},
  \bibinfo{numpages}{12}~pages.
\newblock


\bibitem[Nichols(2021b)]%
        {nichols2021trust}
\bibfield{author}{\bibinfo{person}{Kathleen Nichols}.}
  \bibinfo{year}{2021}\natexlab{b}.
\newblock \showarticletitle{Trust schemas and ICN: key to secure home IoT}. In
  \bibinfo{booktitle}{\emph{Proceedings of the 8th ACM Conference on
  Information-Centric Networking}}. \bibinfo{pages}{95--106}.
\newblock


\bibitem[Pesavento et~al\mbox{.}(2022)]%
        {davide2022pion}
\bibfield{author}{\bibinfo{person}{Davide Pesavento}, \bibinfo{person}{Junxiao
  Shi}, \bibinfo{person}{Kerry McKay}, {and} \bibinfo{person}{Lotfi
  Benmohamed}.} \bibinfo{year}{2022}\natexlab{}.
\newblock \showarticletitle{PION: Password-based IoT Onboarding Over Named Data
  Networking}. In \bibinfo{booktitle}{\emph{2022 IEEE International Conference
  on Communications}}. IEEE.
\newblock


\bibitem[{Pesavento, Davide and Shi, Junxiao}(2022)]%
        {pion-spec}
\bibfield{author}{\bibinfo{person}{{Pesavento, Davide and Shi, Junxiao}}.}
  \bibinfo{year}{2022}\natexlab{}.
\newblock \bibinfo{title}{PION Specification}.
\newblock
\newblock
\urldef\tempurl%
\url{https://github.com/usnistgov/PION/blob/main/docs/protocol.md}
\showURL{%
\tempurl}


\bibitem[Ramani et~al\mbox{.}(2020)]%
        {ramani2020ndnviber}
\bibfield{author}{\bibinfo{person}{Sanjeev~Kaushik Ramani},
  \bibinfo{person}{Proyash Podder}, {and} \bibinfo{person}{Alex Afanasyev}.}
  \bibinfo{year}{2020}\natexlab{}.
\newblock \showarticletitle{NDNViber: Vibration-Assisted Automated
  Bootstrapping of IoT Devices}. In \bibinfo{booktitle}{\emph{2020 IEEE
  International Conference on Communications Workshops (ICC Workshops)}}. IEEE,
  \bibinfo{pages}{1--6}.
\newblock


\bibitem[{The NDN Team}(2022a)]%
        {ndnpythonrepo}
\bibfield{author}{\bibinfo{person}{{The NDN Team}}.}
  \bibinfo{year}{2022}\natexlab{a}.
\newblock \bibinfo{title}{NDN Python Repo}.
\newblock
\newblock
\urldef\tempurl%
\url{https://github.com/UCLA-IRL/ndn-python-repo}
\showURL{%
\tempurl}


\bibitem[{The NDN Team}(2022b)]%
        {ndncert-code}
\bibfield{author}{\bibinfo{person}{{The NDN Team}}.}
  \bibinfo{year}{2022}\natexlab{b}.
\newblock \bibinfo{title}{NDNCERT Codebase}.
\newblock
\newblock
\urldef\tempurl%
\url{https://github.com/named-data/ndncert}
\showURL{%
\tempurl}


\bibitem[Ylonen and Lonvick(2006)]%
        {ylonen2006secure}
\bibfield{author}{\bibinfo{person}{Tatu Ylonen} {and} \bibinfo{person}{Chris
  Lonvick}.} \bibinfo{year}{2006}\natexlab{}.
\newblock \bibinfo{booktitle}{\emph{The secure shell (SSH) transport layer
  protocol}}.
\newblock \bibinfo{type}{{T}echnical {R}eport}. \bibinfo{institution}{RFC 4253,
  January}.
\newblock


\bibitem[Yu et~al\mbox{.}(2021)]%
        {yu2021enabling}
\bibfield{author}{\bibinfo{person}{Tianyuan Yu}, \bibinfo{person}{Philipp
  Moll}, \bibinfo{person}{Zhiyi Zhang}, \bibinfo{person}{Alexander Afanasyev},
  {and} \bibinfo{person}{Lixia Zhang}.} \bibinfo{year}{2021}\natexlab{}.
\newblock \showarticletitle{Enabling Plug-n-Play in Named Data Networking}. In
  \bibinfo{booktitle}{\emph{MILCOM 2021-2021 IEEE Military Communications
  Conference (MILCOM)}}. IEEE, \bibinfo{pages}{562--569}.
\newblock


\bibitem[Yu et~al\mbox{.}(2015)]%
        {yu2015schematizing}
\bibfield{author}{\bibinfo{person}{Yingdi Yu}, \bibinfo{person}{Alexander
  Afanasyev}, \bibinfo{person}{David Clark}, \bibinfo{person}{kc claffy},
  \bibinfo{person}{Van Jacobson}, {and} \bibinfo{person}{Lixia Zhang}.}
  \bibinfo{year}{2015}\natexlab{}.
\newblock \showarticletitle{Schematizing Trust in Named Data Networking}. In
  \bibinfo{booktitle}{\emph{Proceedings of the 2nd ACM Conference on
  Information-Centric Networking}} (San Francisco, California, USA)
  \emph{(\bibinfo{series}{ACM-ICN ’15})}. \bibinfo{publisher}{Association for
  Computing Machinery}, \bibinfo{address}{New York, NY, USA},
  \bibinfo{pages}{177–186}.
\newblock
\showISBNx{9781450338554}


\bibitem[Yu et~al\mbox{.}(2014a)]%
        {yu2014endorsement}
\bibfield{author}{\bibinfo{person}{Yingdi Yu}, \bibinfo{person}{Alexander
  Afanasyev}, \bibinfo{person}{Zhenkai Zhu}, {and} \bibinfo{person}{Lixia
  Zhang}.} \bibinfo{year}{2014}\natexlab{a}.
\newblock \showarticletitle{An endorsement-based key management system for
  decentralized NDN chat application}.
\newblock \bibinfo{journal}{\emph{University of California, Los Angeles, Tech.
  Rep. NDN-0023}} (\bibinfo{year}{2014}).
\newblock


\bibitem[Yu et~al\mbox{.}(2014b)]%
        {yu2014ndn}
\bibfield{author}{\bibinfo{person}{Yingdi Yu}, \bibinfo{person}{A Afanasyev},
  \bibinfo{person}{Z Zhu}, {and} \bibinfo{person}{L Zhang}.}
  \bibinfo{year}{2014}\natexlab{b}.
\newblock \showarticletitle{Ndn technical memo: Naming conventions}.
\newblock \bibinfo{journal}{\emph{NDN, NDN Memo, Technical Report NDN-0023}}
  (\bibinfo{year}{2014}).
\newblock


\bibitem[Zhang et~al\mbox{.}(2018a)]%
        {zhang2018hostmodel}
\bibfield{author}{\bibinfo{person}{Haitao Zhang}, \bibinfo{person}{Yanbiao Li},
  \bibinfo{person}{Zhiyi Zhang}, \bibinfo{person}{Alexander Afanasyev}, {and}
  \bibinfo{person}{Lixia Zhang}.} \bibinfo{year}{2018}\natexlab{a}.
\newblock \showarticletitle{NDN Host Model}.
\newblock \bibinfo{journal}{\emph{SIGCOMM Comput. Commun. Rev.}}
  \bibinfo{volume}{48}, \bibinfo{number}{3} (\bibinfo{date}{Sept.}
  \bibinfo{year}{2018}), \bibinfo{pages}{35–41}.
\newblock


\bibitem[Zhang et~al\mbox{.}(2016)]%
        {zhang2016sharing}
\bibfield{author}{\bibinfo{person}{Haitao Zhang}, \bibinfo{person}{Zhehao
  Wang}, \bibinfo{person}{Christopher Scherb}, \bibinfo{person}{Claudio
  Marxer}, \bibinfo{person}{Jeff Burke}, \bibinfo{person}{Lixia Zhang}, {and}
  \bibinfo{person}{Christian Tschudin}.} \bibinfo{year}{2016}\natexlab{}.
\newblock \showarticletitle{Sharing mhealth data via named data networking}. In
  \bibinfo{booktitle}{\emph{Proceedings of the 3rd ACM Conference on
  Information-Centric Networking}}. \bibinfo{pages}{142--147}.
\newblock


\bibitem[Zhang et~al\mbox{.}(2014)]%
        {zhang2014ndn}
\bibfield{author}{\bibinfo{person}{Lixia Zhang}, \bibinfo{person}{Alexander
  Afanasyev}, \bibinfo{person}{Jeffrey Burke}, \bibinfo{person}{Van Jacobson},
  \bibinfo{person}{kc claffy}, \bibinfo{person}{Patrick Crowley},
  \bibinfo{person}{Christos Papadopoulos}, \bibinfo{person}{Lan Wang}, {and}
  \bibinfo{person}{Beichuan Zhang}.} \bibinfo{year}{2014}\natexlab{}.
\newblock \showarticletitle{Named Data Networking}.
\newblock \bibinfo{journal}{\emph{SIGCOMM Comput. Commun. Rev.}}
  \bibinfo{volume}{44}, \bibinfo{number}{3} (\bibinfo{date}{July}
  \bibinfo{year}{2014}), \bibinfo{pages}{66–73}.
\newblock
\showISSN{0146-4833}


\bibitem[Zhang et~al\mbox{.}(2020)]%
        {zhang2020certificate}
\bibfield{author}{\bibinfo{person}{Zhiyi Zhang}, \bibinfo{person}{Su~Yong
  Wong}, \bibinfo{person}{Junxiao Shi}, \bibinfo{person}{Davide Pesavento},
  \bibinfo{person}{Alexander Afanasyev}, {and} \bibinfo{person}{Lixia Zhang}.}
  \bibinfo{year}{2020}\natexlab{}.
\newblock \showarticletitle{On Certificate Management in Named Data
  Networking}.
\newblock \bibinfo{journal}{\emph{arXiv preprint arXiv:2009.09339}}
  (\bibinfo{year}{2020}).
\newblock


\bibitem[Zhang et~al\mbox{.}(2022)]%
        {zhang2022sovereign}
\bibfield{author}{\bibinfo{person}{Zhiyi Zhang}, \bibinfo{person}{Tianyuan Yu},
  \bibinfo{person}{Xinyu Ma}, \bibinfo{person}{Yu Guan},
  \bibinfo{person}{Philipp Moll}, {and} \bibinfo{person}{Lixia Zhang}.}
  \bibinfo{year}{2022}\natexlab{}.
\newblock \showarticletitle{Sovereign: Self-contained Smart Home with
  Data-centric Network and Security}.
\newblock \bibinfo{journal}{\emph{IEEE Internet of Things Journal}}
  (\bibinfo{year}{2022}).
\newblock


\bibitem[Zhang et~al\mbox{.}(2018b)]%
        {zhang2018nac}
\bibfield{author}{\bibinfo{person}{Zhiyi Zhang}, \bibinfo{person}{Yingdi Yu},
  \bibinfo{person}{Sanjeev~Kaushik Ramani}, \bibinfo{person}{Alex Afanasyev},
  {and} \bibinfo{person}{Lixia Zhang}.} \bibinfo{year}{2018}\natexlab{b}.
\newblock \showarticletitle{NAC: Automating access control via Named Data}. In
  \bibinfo{booktitle}{\emph{MILCOM 2018-2018 IEEE Military Communications
  Conference (MILCOM)}}. IEEE, \bibinfo{pages}{626--633}.
\newblock


\bibitem[Zhang et~al\mbox{.}(2018c)]%
        {zhang2018security}
\bibfield{author}{\bibinfo{person}{Zhiyi Zhang}, \bibinfo{person}{Yingdi Yu},
  \bibinfo{person}{Haitao Zhang}, \bibinfo{person}{Eric Newberry},
  \bibinfo{person}{Spyridon Mastorakis}, \bibinfo{person}{Yanbiao Li},
  \bibinfo{person}{Alexander Afanasyev}, {and} \bibinfo{person}{Lixia Zhang}.}
  \bibinfo{year}{2018}\natexlab{c}.
\newblock \showarticletitle{An overview of security support in named data
  networking}.
\newblock \bibinfo{journal}{\emph{IEEE Communications Magazine}}
  \bibinfo{volume}{56}, \bibinfo{number}{11} (\bibinfo{year}{2018}),
  \bibinfo{pages}{62--68}.
\newblock


\bibitem[{Zhang, Zhiyi and Shi, Junxiao and Pesavento, Davide}(2022)]%
        {ndncert-spec}
\bibfield{author}{\bibinfo{person}{{Zhang, Zhiyi and Shi, Junxiao and
  Pesavento, Davide}}.} \bibinfo{year}{2022}\natexlab{}.
\newblock \bibinfo{title}{NDNCERT v0.3 Specification}.
\newblock
\newblock
\urldef\tempurl%
\url{https://github.com/named-data/ndncert/wiki/NDNCERT-Protocol-0.3}
\showURL{%
\tempurl}


\end{thebibliography}
